**TITLE**

Enhanced Knee Kinematics: Leveraging Deep Learning and Morphing Algorithms for 3D Implant Modeling

**AUTHORS**

Viet Dung Nguyen [1], Ph.D.

Michael T. LaCour [1,2], Ph.D.

Richard D. Komistek [2], Ph.D.

[1] Orthopaedic Innovation Technology Center

[2] University of Tennessee at Knoxville

**CORRESPONDING AUTHOR**

Viet Dung Nguyen

Orthopaedic Innovation Technology Center

Knoxville, TN 37916

vdnguyen@oitctn.com

**WORD COUNT**

Abstract : 330

Main text (Excluded references) : 2500

**ABSTRACT**

In the realm of orthopedic surgery and biomedical engineering, the accurate reconstruction of implanted knee models plays a pivotal role in enhancing preoperative planning, optimizing implant design, and improving surgical outcomes. Traditional methods of constructing these models often rely on manual segmentation techniques, which are labor-intensive and prone to human error. To address these challenges, this study proposes a novel approach that leverages machine learning (ML) algorithms and morphing techniques for the precise reconstruction of 3D implanted knee models.

The proposed methodology begins with the acquisition of preoperative imaging data, such as four fluoroscopy or x-ray images of the patient's knee joint. Subsequently, a convolutional neural network (CNN) is trained to automatically segment the femur contour of the implanted components. This automated segmentation process significantly reduces the time and effort required for manual delineation while ensuring high accuracy and reliability.

Following the segmentation stage, a morphing algorithm is employed to generate a personalized 3D model of the implanted knee joint. This algorithm utilizes segmented data as well as biomechanical principles to simulate the shape of the knee joint, considering factors such as implant position, size, and orientation. By integrating morphological data with implant-specific parameters, the reconstructed models reflect the patient's implant anatomy and implant configuration.

The proposed approach's effectiveness is demonstrated through thorough quantitative evaluations, including comparisons with ground truth data and existing reconstruction techniques. Across 19 test cases involving various implant types like cruciate retaining and posterior stabilized TKAs, the comparison with known ground truths shows that the ML-based segmentation method achieves superior accuracy and consistency compared to manual segmentation approaches, with an average RMS error of 0.58±0.14 mm.

Overall, this research contributes to advancing the field of orthopedic surgery by providing a robust framework for the automated reconstruction of implanted knee models. By harnessing the power of machine learning and morphing algorithms, clinicians and researchers can gain valuable insights into patient-specific knee anatomy, implant biomechanics, and surgical planning, ultimately leading to improved patient outcomes and enhanced quality of care.

**1 Introduction**

In the field of orthopedic surgery and biomedical engineering, understanding kinematics in total knee arthroplasty (TKA) is crucial for optimizing surgical outcomes and ensuring proper joint function and longevity [1]. A 3-dimensional (3D) model is essential for accurately simulating and analyzing the intricate kinematics of the knee joint during total knee arthroplasty, aiding in precise surgical planning and implant placement [2]. Without this 3D model, called a CAD model, it's nearly impossible to compute the implant kinematics [3].

A 3D implant model can be reconstructed by the conventional approaches such as computed-tomography (CT) or magnetic-resonance-imaging (MRI) scans that are accurate but expensive [4]. Current methods for 3D implant reconstruction encompass a spectrum of techniques, ranging from manual segmentation to advanced machine learning-based approaches. Traditional methods involve laborious manual delineation of implant boundaries on medical images, followed by manual reconstruction to generate 3D models. While providing a high level of control, these methods are labor-intensive [5], time-consuming [6], and susceptible to inter-observer variability, leading to inconsistency and potential errors in the reconstructed models. Semi-automated approaches attempt to mitigate these drawbacks by combining manual initialization with automated algorithms for segmenting and reconstructing implant geometries. However, they still require significant user intervention and may suffer from accuracy issues, particularly in cases of complex implant geometry or image artifacts.

In recent years, there has been a growing interest in image-based reconstruction methods, leveraging image processing algorithms to automatically extract implant contours from medical images. While promising in terms of automation and efficiency, these methods are often sensitive to image noise, artifacts, and variations in implant appearance, which can affect the accuracy of the reconstructed models. Additionally, machine learning-based techniques, particularly deep learning architectures like convolutional neural networks (CNNs), have shown remarkable capabilities in image segmentation and feature extraction tasks. However, challenges such as the need for large, annotated datasets, model generalization across different implant types, and interpretability of the learned representations remain significant drawbacks of ML-based approaches.

To overcome these limitations, this study introduces a fast and fully automated approach that utilizes only four fluoroscopic images, as seen in Figure 1, by using machine learning (ML) algorithm and morphing technique to achieve accurate reconstructions of 3D implanted knee models with lower radiation exposure than CT scan. The task starts with extracting the two-dimensional (2D) contours of implants from fluoroscopic images via image segmentation. To obviate the need for manual intervention in this segmentation process, a deep neural network model was trained and deployed for automating the image segmentation task [7]. Subsequently, a morphing algorithm [8] is integrated with the aforementioned machine learning framework to facilitate the complete automatic reconstruction of three-dimensional (3D) implant models.

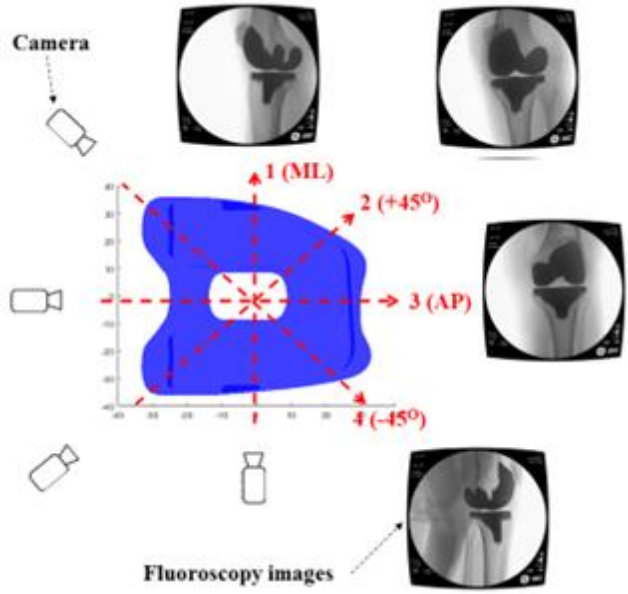

Figure 1: Reconstruction of an implanted knee by using four fluoroscopic images

**2 Method**

2.1 Study design

In this study, we revolutionize segmentation practices through the power of deep learning. Our approach streamlines the process by eliminating manual segmentation entirely. We prepare our data for training our deep neural network, utilizing a combination of fluoroscopic images and CAD models. For the fluoroscopic images of Total Knee Arthroplasty (TKA) patients, our data preparation involves an exhaustive analysis encompassing 5 comprehensive studies [9, 3, 1, 10, 11]. These studies collectively incorporate 939 patients of 6,000 fluoroscopic images, encompassing a diverse range of implant types such as PS and CR. From this vast dataset, we meticulously extract high-quality images, overlaying them with corresponding CAD models to ensure precision.

These refined images are then utilized to validate the accuracy of our patient output with remarkable success. As for the CAD models, we source ground truths directly from the design company, ensuring the reliability and authenticity of our data. This approach guarantees the robustness and effectiveness of our methodology in advancing the field of medical imaging.

In this study, our reconstruction algorithm leverages data from 19 patients. Each patient contributes four distinct fluoroscopic images captured at various angles: anterior-posterior (AP), medial-lateral (ML), and rotations of ±45 degrees. This comprehensive approach ensures that we capture the richest possible information from each patient. Our research aims to reconstruct knee implant models for all 19 patients based on their corresponding fluoroscopic images. This endeavor promises to yield invaluable insights into implant positioning and performance, thereby contributing significantly to the field of orthopedic surgery.

2.2 Image segmentation and edge detection

With the above fluoroscopic images and CAD models, a validated 3D-2D registration technique [12] matches the projection of the 3D model with the silhouette of the implant in the 2D image. By using the voxelization projection [13], the known registration transformations can produce segmented images of both femur and tibia directly from the registered models. This process generates a raw segmentation dataset consisting of 6,000 pairs of the original fluoroscopic images and corresponding segmented and labeled images of either femur or tibia. From here, by converting this dataset to the COCO (Common Objects in Context) format [14], which is standard for image processing, it is possible to not only train an appropriate leaning model but also quantifiably compare and verify the accuracy and performance of the proposed deep learning model against other existing methods.

Among advanced deep neural networks available for this task, such as FCN [15], CRF-RNN [16], DPN [17], GCN [18], and Yolact [19], the Yolact algorithm was chosen for its high accuracy and speed, making it a logical choice for knee image segmentation. The deep neural model based Yolact takes original fluoroscopic images as inputs and generates three layers of outputs representing labels, boxes, and masks (as depicted in Figure 2). The 6,000 implanted knee images were separated into a training set (90%) and testing set (10%). The training process employs the backpropagation method, enhanced by GPU acceleration for faster computation. This procedure is performed on the Alienware Aurora R13 Desktop, equipped with 32 Gb RAM and NVIDIA GeForce RTX 3080 GPU. It runs for 1 million epochs with a batch size of 8 and takes approximately 3 days to complete the training process. This training step is completed in Python 3.7. After getting the segmented image, the edge is detected by the contour of the mask image of the femur with one pixel thickness as in Figure 2. The edge detection step is completed in Matlab.

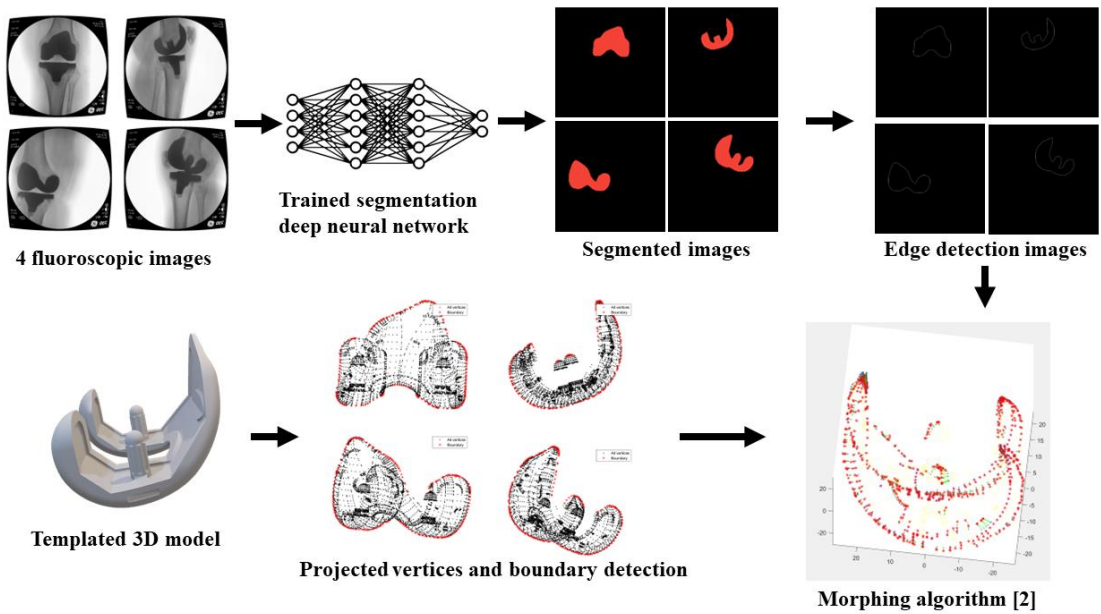

Figure 2: Method of fully automated reconstruction using deep learning and morphing algorithm

2.3 Morphing algorithm

Following the segmentation stage, a morphing algorithm is employed to generate a personalized 3D model of the implanted knee joint. This algorithm utilizes segmented data as well as biomechanical principles to simulate the shape of the knee joint, considering factors such as implant position, size, and orientation. By integrating morphological data with implant-specific parameters, the reconstructed models reflect the patient's implant anatomy and implant configuration. The algorithm of reconstructing a CAD model given the intial and target contours is published at [8]. This step is completed in Matlab.

## 3 Computational Results

For the image segmentation accuracy, the simulation results of the trained deep learning model revealed good accuracy via comparison of the predicted segmentation images with the ground truth of 0.89 mAP, as seen in two examples of PS and CR implants in Figure 3. For optimal reconstruction accuracy, two comprehensive algorithm evaluation studies were devised to assess the precision of morphed models. Initially, the algorithm underwent a process of morphing into a smaller TKA template component model, which was subsequently juxtaposed against various larger, established sizes. This evaluation was conducted within the framework where the genuine larger CAD models were readily available for scrutiny, enabling a direct comparison with the resultant models generated by the morphing algorithm. Precisely, the alignment of the morphed model and the genuine model was achieved through ICP, following which the error of each vertex of the morphed model was meticulously calculated based on its distance from the surface of the genuine model. The efficacy of the proposed algorithm was scrutinized by analyzing two distinct types of errors: (i) the root-mean-square (RMS) error, and (ii) the largest source error among all vertices.

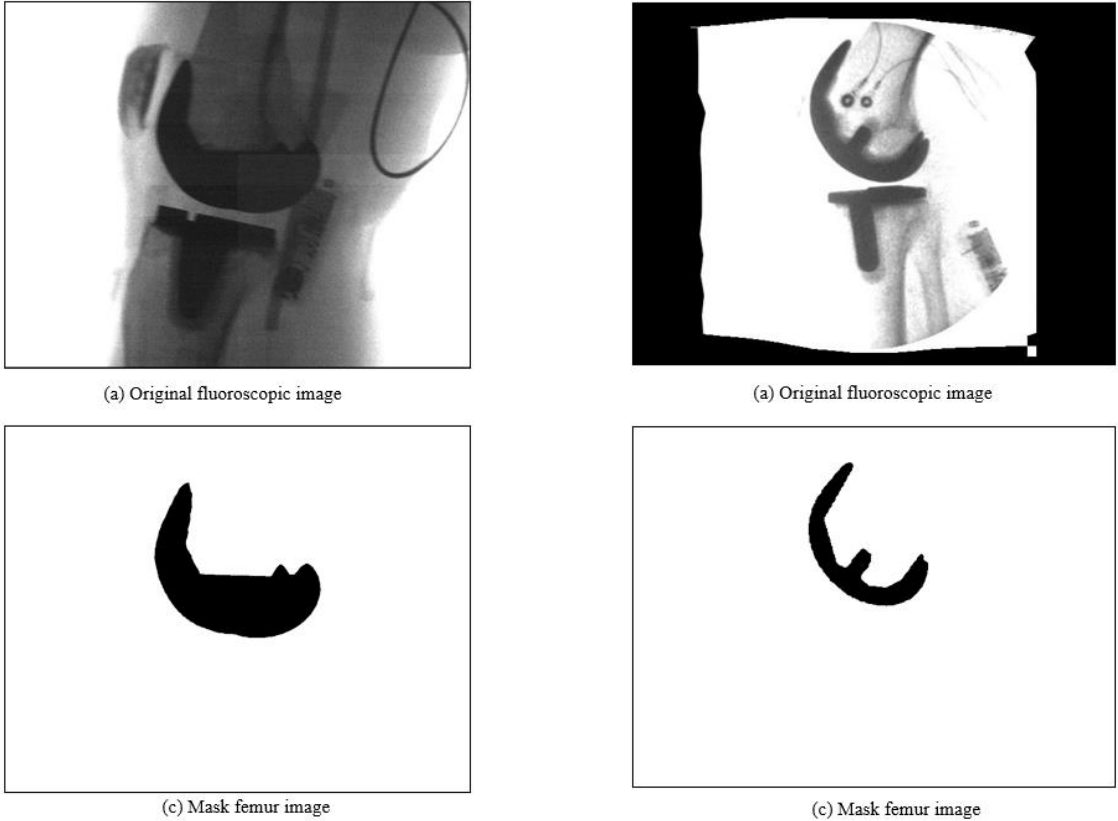

Figure 3: A sample of an original fluoroscopy image of the PS and CR implant patients and the corresponding segmented images by deep learning.

In Figure 4, a heatmap presents a quantitative comparison between the sample morphed model and the established target CAD model. In this specific instance, the root-mean-square (RMS) error amounted to 0.43 mm, with the absolute largest source of error registering at 2.21 mm. Encouragingly, the most significant error was localized to the internal surfaces of the model, as depicted in Figure 4 (right), an area typically less relevant for fluoroscopic assessments, which prioritize the outer articulating surfaces of the implants. Moreover, the articulating surface of the J-curve exhibits notably diminished errors, ranging from -0.2 mm to 0.1 mm, affirming the accuracy of the articulating geometry, as showcased in Figure 4 (left).

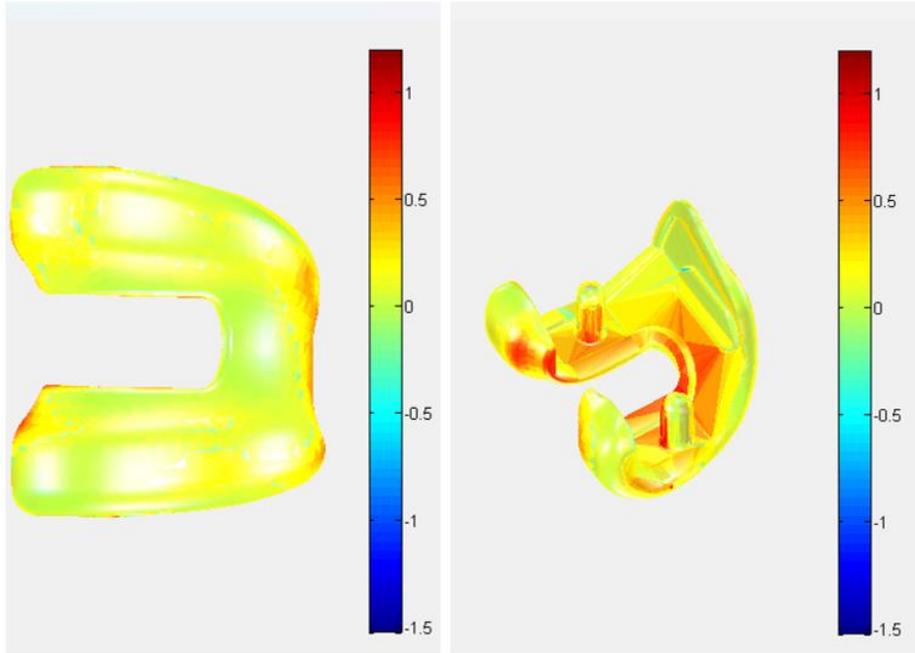

Figure 4: Heatmap showing the comparison between the reconstructed model and the ground truth

In this study, the morphing algorithm was employed to reconstruct 19 implant CAD models of varying sizes sourced from different companies. The fidelity of these morphed models compared to ground truth data, irrespective of implant type, is meticulously detailed in Table 1, showcasing RMS error and largest source error metrics. The reconstructed models exhibit an impressive average RMS error of $0.58 \pm 0.14$ mm, ranging from 0.39 to 0.84 mm. Similarly, the average of the largest errors stands at $2.77 \pm 0.79$ mm, occurring predominantly on internal implant features, within a range of 1.54 to 4.52 mm.

Table 1: The evaluation of 19 reconstruction models with cruciate retaining and posterior stabilized TKA reconstructions evaluations

| Test Cases | 1 | 2 | 3 | 4 | 5 | 6 | 7 |
|---|---|---|---|---|---|---|---|
| RMS Error (mm) | 0.46 | 0.54 | 0.58 | 0.84 | 0.45 | 0.66 | 0.39 |
| Largest Error (mm) | 1.54 | 2.75 | 1.76 | 3.68 | 3.27 | 3.47 | 1.94 |
| **Test Cases** | **8** | **9** | **10** | **11** | **12** | **13** | **14** |
| RMS Error (mm) | 0.53 | 0.49 | 0.62 | 0.49 | 0.81 | 0.65 | 0.53 |
| Largest Error (mm) | 1.86 | 4.52 | 2.68 | 2.71 | 2.97 | 2.97 | 2.74 |
| **Test Cases** | **15** | **16** | **18** | **18** | **19** | **Average (All)** | |
| RMS Error (mm) | 0.45 | 0.51 | 0.83 | 0.56 | 0.81 | 0.58 ± 0.14 | |
| Largest Error (mm) | 1.75 | 2.52 | 3.48 | 2.35 | 3.75 | 2.77 ± 0.79 | |

For kinematic evaluation, Figure 5 vividly illustrates the outputs using the reconstructed models alongside the known models, as a subject engages in a deep knee bend activity. Remarkably, the reconstructed models mirror kinematic outputs that closely parallel ground truths, both visually and quantitatively. In numerical terms, the average error in femorotibial translation between the reconstructed approach and ground truth registers at 0.63 ± 0.39 mm, while for axial rotation, it stands at 0.92 ± 0.42 degrees. These errors, although present, fall well within an acceptable range for kinematic evaluation. Notably, they compare favorably with typical manual registration errors of approximately 0.65 mm for translation and 1.5 degrees for rotation [4], underscoring the reliability of the morphing algorithm in this critical domain.

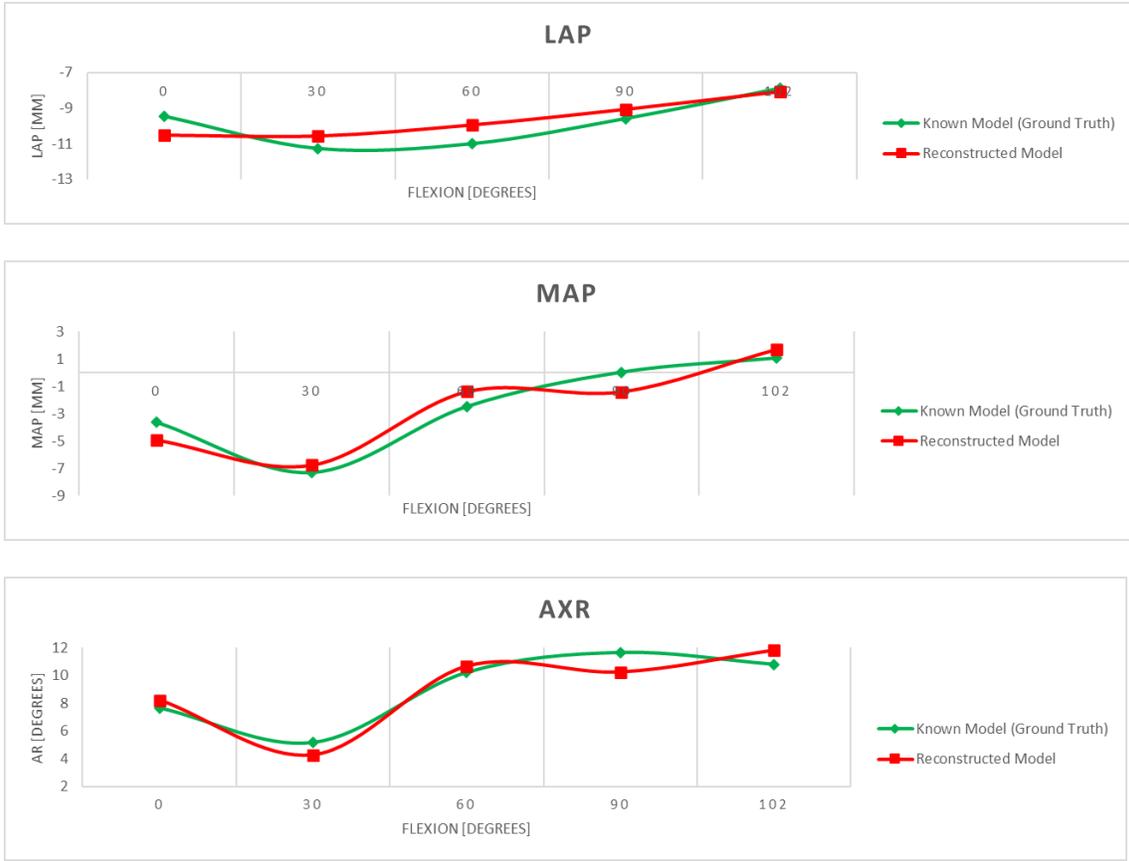

Figure 5: Kinematics comparison between the reconstructed model and the ground truth

**4 Discussion**

One of the key points for discussion is the effectiveness of integrating machine learning and morphing algorithms in reconstructing 3D models of implanted knees. The results have shown promising outcomes, with the combined approach offering superior accuracy compared to traditional methods. This highlights the potential of leveraging advanced computational techniques to overcome challenges in orthopedic imaging and modeling.

Additionally, the discussion explores the practical applications of the reconstructed knee models. These models can serve as valuable tools for preoperative planning, allowing surgeons to visualize and analyze the anatomical structures and implant placements with high precision. Moreover, they can facilitate the development and testing of new implant designs, leading to improved outcomes for patients undergoing knee replacement surgeries. In contrast to CT scans, which involve higher radiation exposure, this method offers a safer alternative by utilizing only four x-ray images. The reduced number of x-rays significantly mitigates radiation risks for patients undergoing knee reconstruction procedures. This not only enhances the overall safety profile of the imaging process but also minimizes potential long-term health concerns associated with repeated exposure to ionizing radiation. Consequently, the utilization of this method presents a compelling advantage in orthopedic imaging, prioritizing patient well-being without compromising the quality or accuracy of the reconstructed knee models.

The method presented in this study not only boasts increased accuracy in specific cases but also offers practical advantages in terms of efficiency and cost-effectiveness. With reduced processing time and lower labor costs compared to traditional techniques, this approach holds promise for widespread adoption in medical settings. Furthermore, its potential for automation makes it particularly well-suited for use in emergency rooms, where rapid and accurate assessments are crucial. By streamlining the process of reconstructing knee models, this method has the potential to revolutionize orthopedic care, providing timely and reliable support for medical professionals in critical decision-making scenarios.

However, it's important to acknowledge the limitations of the study. Despite the advancements achieved, there may still be challenges in accurately capturing certain nuances of knee anatomy or implant variations. Furthermore, the computational resources required for implementing these algorithms may pose constraints, particularly in clinical settings with limited access to high-performance computing resources. While not consistently more accurate on average, the methodology showcased in this study exhibits enhanced precision in specific cases where traditional approaches falter due to missing portions in images, as shown in Figure 6. This advantage underscores the significance of integrated machine learning and morphing algorithms, which excel in compensating for gaps or incomplete data, resulting in superior reconstructions. By addressing these inherent limitations, the proposed approach offers a tailored solution that can significantly improve the overall reliability and utility of reconstructed knee models, particularly in scenarios where conventional methods struggle to provide comprehensive representations.

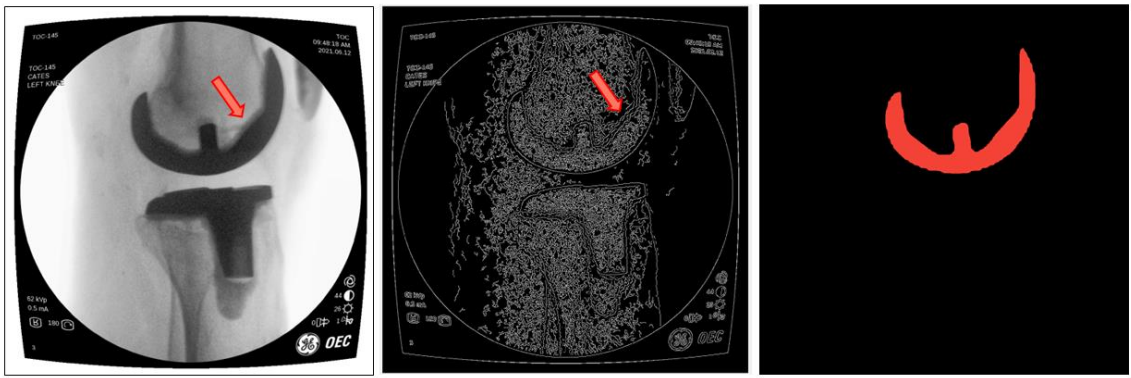

Figure 6: An example of fluoroscopic image (lef) were manual segmentation failure (middle) vs. deep learning success (right).

Looking ahead, future research directions could focus on addressing these limitations and expanding the applicability of the proposed methodology. This may involve refining the algorithms to handle a wider range of implant types and patient anatomies, as well as exploring ways to streamline the reconstruction process for real-time or near-real-time applications. Additionally, investigating the potential integration of other imaging modalities or data sources could further enhance the accuracy and comprehensiveness of the reconstructed models. Thus, the discussion section emphasizes the transformative potential of combining machine learning and morphing algorithms for 3D reconstruction of implanted knee models. While there are challenges to overcome and avenues for further research, the findings of this study pave the way for advancements in orthopedic surgery planning and patient care.

## 5 Conclusion

In conclusion, the integration of machine learning and morphing algorithms has proven to be a powerful approach for reconstructing 3D models of implanted knees. Through the utilization of advanced computational techniques, this paper has demonstrated significant advancements in accurately capturing the complex geometry and structure of knee implants. The synergistic combination of machine learning's ability to learn from data patterns and morphing algorithms' capacity to adaptively deform shapes has enabled enhanced reconstruction outcomes, offering valuable insights for medical professionals and researchers alike. Moving forward, further exploration and refinement of these methodologies hold promise for continued advancements in orthopedic surgery planning, implant design optimization, and patient-specific treatment strategies.

**Acknowledgement**

# List of References